# ALBERT EINSTEIN - A PIOUS ATHEIST

*We consider Einstein's attitude with regard to religion
both from sociological and epistemological points of view. An attempt
to put it into a wider socio-historical perspective has been made,
with the emphasis on his ethnic and religious background. The great
scientist was neither anatheist nor a believer in the orthodox sense
and the closest labels one might apply would be pantheism/cosmism
(ontological view) and agnosticism (epistemological view). His
ideas on the divine could be considered as a continuation of a line
that can be traced back to Philo of Alexandria, who himself followed
the Greek Stoics and Neoplatonists and especially Baruch Spinoza.
Einstein's scientific (or rational) and religious (or intuitive)
thinking was deeply rooted in the Hellenic culture.*

## 1. Prologue

As the centenary of Einstein's annus mirabilis evolved, one of the most
notable personalities of the previous century became the focus of the
world's attention. Many aspects of his extraordinary personality were
subjected to scrutiny, including his relation to religion. Although
Einstein was not a professional religious thinker, his fame and authority
in pure science made his personal beliefs both interesting and influential.
Was Albert Einstein a religious man? But before we attempt to answer this
question, another one seems in order. Is this a proper question at all?
Can such a profound scientific mind be analyzed in less profound terms?
In particular, can the mind that has transformed our most fundamental
concepts, such as space, time and causality, put into standard modes of
thought the most elusive notions like faith or God(s) for example? We
shall arguein the following that (i) Einstein was arguing for a new
kind of religion and (ii) at the same time was playing a role, albeit
subconsciously, of a prophet, even a god.

## 2. Hellenism and Judaism

When Albert was twelve, a young student from Poland who used to come
to Einstein's home for dinner, brought him a popular edition of Euclid's
Elements. It seems to have been a decisive moment for the young boy,
who was at the time obsessed by the biblical fables. He must have
already felt a suspicion about the historical reality of biblical
events and the issue of their truth must have been raised in his mind,
when he encountered Euclid's work. Unlike biblical authors who offered
incredible fables and interpretations, Albert found in the Elements the absolute (mathematical) truth, exposed by an iron logic, firm and undeniable. Incidentally, the treatise was composed in the same period (3rd century BC) in which the translation of Holy Scriptures (Septuagint) was made. Both books epitomized the Judaic and Hellenistic ideologies, as the paradigms of the fictitious and the rational. The further intellectual development of the young Albert was determined by the interplay of his Jewish ethnic origin and rational Hellenistic education.

Technically speaking, Einstein remained within the sphere of the Judaic tradition. But his personal development led him to depart from Orthodox Judaism, and this evolution resulted in two main accomplishments. Firstly, this started from the traditional common religion as practised by clergymen and evolved to a progressively more abstract concept of piety; secondly, he found in retrospect that mankind followed the same route. One might consider this sort of individual and historical correlation as another manifestation of the famous Hoeckel's thesis: Ontogeny is a recapitulation of phylogeny. This development was by no means original, and could be traced in at least two Western traditions: Hellenic and Judaic. Since these two traditions turn out to be crucial for understanding Einstein's view of religious phenomena, we now sketch their common features and differences of their content and ultimate historical fates.

Both traditions followed the same evolutionary pattern, starting from a common religion, more precisely - religions designed for common people. The principal features of these stages are personalized, even anthropic gods.  The Greeks had a well-developed, though not unique and fixed Pantheon, whereas the early Hebrews started with a number of henotheistic gods as well. But after this initial phase, these two traditions diverged. Greek religion remained polytheistic until the appearance of Christianity [1] as a quasi-monotheistic faith, whereas the Hebrews soon reduced their concept of divine oligarchy to the monotheistic one, their own tribal god, (Jehovah). Later on this god became the God, a unique deity, which acquired two principal, albeit contradictory, attributes. He was the God of all mankind, the universal Demiurge, and also the tribal Jewish god bound by covenant to his chosen people. This dichotomy was resolved by Christian teaching, which resulted in the rapid spreading of the Judaic mythology over a large part of the globe. But apart from her Christian "heresy" the Judaic common religion has retained its principal features.

We now turn to the intellectual religious sphere, that of thinkers and philosophers. Greek development on this more abstract level followed the line of Xenophanes [2], Plato and Aristotle, who conceived an abstract deity, devoid of banal anthropic properties and disinterested in human affairs. To Aristotle the First Mover was necessary just to start the life of Nature, and whose further history was to be governed by the laws of Nature which were to be inferred by human mind. A particular concept of the supernatural Demiurge was conceived by Anaxagoras [2], who introduced the notion of the Mind (Νους) engaged in creating the Cosmos out of a primordial mixture of seeds (σπερματα). It seems that Anaxagoras' Mind was something between a moving agency and a natural principle. In any case, this Greek philosophical line was interrupted by the Christian faith and was never canonized into a common religion.

Judaic tradition followed two main streams. One was an orthodox rabinical one, mainly present today within the Jewish milieu, the other adopted an esoteric route, developing somewhat extravagant ideas on divinity, as is the case with Kabbala. Extravagant as they appear, some of the esoteric concepts resemble modern cosmological scenarios of the World creation remarkably well. A case in point is the Kabbalistic constructs of *Ain Soph* and *Zimzum* [3], with its striking similarity to the Big Bang inflationary paradigm. The philosophy of Baruch Spinoza lies somewhat between the orthodox and the esoteric and deserves our particular attention here.

## 3. Spinoza and pantheism

Broadly speaking, Spinoza's doctrine may be considered as a continuation of the Judaic philosophical tradition, whose beginnings can be traced back to Philo of Alexandria [4,5]. Living in the metropolis of the Hellenistic world, with a large Greek population and a numerous Jewish community, Philo's concern was mainly in reconciling the Jewish faith based on the Torah mythology with the superior Greek philosophical teachings. He was squeezed between the elaborate rational systems of Plato, Aristotle and Neopythagoreans on one side, and the sclerotic canonized religious dogmas and tradition of the Holy Scripts which were not to be modified. Therefore Philo resorted to an allegorical interpretation of the written sacrosanct texts. One of his first tasks was to depersonalize God, interpreting the assertion that God created man in his own image as not referring to a physical appearance of the Creator, but to his ethical essence. This departure from the literal meaning was a big step from the common faith toward a more rational, abstract religion, making it a philosophical subject. It is this new concept of God that Spinoza took over and developed further to extreme rational and logical limits.

Spinoza was Einstein's religious hero and for good reason. His teaching epitomized the best amalgamation of the two principal roots on which European culture rested: Hellenistic rationality and Judaic faith. His principal philosophical tract, Ethics [6] had the form of Euclid's *Elements* with its strict deductive structure. Adopting the rational method of exposing his metaphysical ideas, Spinoza adopted at the same time the Greek rationale for scientific truth. Something is not just true; it is so because it must be so, within the context of the overall system. That the Jewish-Dutch philosopher gave the title *Ethics* to his book on the divine nature reveals another rationale for his endeavour - the moral content of the Judaic religion which might be expressed as the relation 'God is ethics'. If the first rationale may be considered as a form of *causa efficientis* within its deductive procedure, the second rationale is manifestly of the nature of *causa finalis*. The ultimate goal is self-divination, immersion into the divine. God is everything man can conceive of, God is Nature. Man is alone before God, since he is a part of Nature. By acquiring the supreme Good, as the essence of divine being, he attains the nature of God. Spinoza's concept of religion is considered as pantheism. The Amsterdam Jewish community interpreted it, rightly, as a form of atheism and banished Spinoza from their community. To state that God is everything is tantamount to saying he is nothing (les extremes se touchent). Pantheism is alien to the European culture and it is more appropriate to relate it to Buddhism, which is not based on the concept of God at all. Judaic tradition might tolerate some form of panentheism, but not pantheism. The former states that everything is God, but God is not everything, he comprises Nature, but the latter does not exhaust his existence. The panentheistic formula thus reads: Everything is in God. By removing all anthropic attributes from the divine, Spinoza dissolved God into cosmic reality and thus annihilated it. Hence his image of God appears closer to that of Anaxagoras' Mind than to Mosaic monotheism.

### 4. Einstein and scientific inquiry

*Va yomer Elohim, yehi or va yehi or.*

Einstein was from early youth inclined to question the unquestionable, suspect the self-evident and test the trivial. His ideas on space (commensurability) and time (simultaneity) put into the formulae that were to be called 'The Special Theory of Relativity', were the fruit of some five years of meditation, as recognized by Einstein himself. In a conversation with Levi-Civita, Einstein remarked that he had had just a couple of ideas in his scientific career. This is essentially true. Einstein's most fundamental contributions to the physical sciences were focussed on a single topic, that of understanding the nature of light. It is well known, according to the testimony of Einstein himself, how the young Albert occupied himself by running in front of a light beam. His enquiries resulted in The Theory of Relativity, which was an answer to two problems. What happens to our comprehension of space and time if the speed of light is absolutely the same for any observer and light motion can not be accelerated nor decelerated? Another important problem where light played a prominent role in Einstein's research was that of the quantum of light, later to be called a photon, which was central to his model of the interaction of the electromagnetic field and inertial matter, as elaborated in the photoelectric effect. It was for this achievement that Einstein received his Nobel Prize in 1921. By taking the idea of a quantum of energy as a particle and representing it as a wave associated with a massive particle, Luis de Broglie opened the route for Erwin Schrödinger to formulate Wave Mechanics, another major achievement of the twenty-century physics.

A great number of Einstein's other valuable contributions may be ascribed to his fascination with light phenomena. The theory of radiation (Einstein coefficients), Bose-Einstein statistics, and even the famous conundrum contrived in the Einstein-Podolsky-Rosen paradox (which has been, perhaps, the most powerful exposition of the weird nature of the Quantum Mechanics) - all refer in one way or another, to electromagnetic field phenomena. In his later years Einstein used to say that all his life he had strived to comprehend the nature of light, and despite some oversimplification this was not far from the truth. Here one encounters one of Einstein's most mysterious features - affinity to mystery.

### 5. Mysterious Einstein versus mysterious Cosmos

Strictly speaking science is not a creative human activity, unlike technique, music or modern art for instance. Science reveals, technique creates. But the further one goes from the ordinary scientific level toward the fundamental issues, the nature of the scientific discovery becomes less of a revelation and more of a creative endeavour. When Einstein set out to apply the formalism of General Relativity to describe the entire Universe, he ceased to purely discover facts and tackled a more ambitious task - to create a picture of the overall physical reality of the Cosmos. Note that in Biblical terms Cosmos is not a purely given entity, it is a Creation. In the prehistoric phase of the mental evolution of *homo sapiens* (the era of the Magi), to name meant to control. This magical ritual was also recorded in the Bible, with Adam given the right to ascribe names to living creatures whose master he was supposed to become. In a more advanced phase naming was not enough and a more detailed description of an entity implied control over it. Knowledge meant power over things. And it is this aspect of understanding which was the rationale for God's forbidding Adam to eat the fruit from the tree of life in Eden, for this allegorical narrative was a story about control. It is the first record of the eternal struggle between the religious and the rational, between the concepts governed by logic and those controlled by fear and mystery.

It was Vico who noted that one best comprehends a concept by inventing it himself [7]. But in the case of Einstein's contributions to what is known today as Special Relativity is not a simple one. As we know today, the ideas were already in the air at the turn of the century and other researchers were on the track, notably Henri Poincare, who first defined the relativity principle [8]. The same holds for the famous formula $E = mc^2$, derived by a number of people before and after 1905. Again, Poincare asserted in 1900 that electromagnetic energy is endowed with an equivalent mass $E/c^2$, but did not pursue the idea to its logical conclusion. It was found many years afterwards that Einstein's original derivation in 1905 was flawed (*circulus vicious*) [9], but he derived it again the following year, this time correctly. It was Max Planck who in 1907, following the original idea due to Hasenöhrl, who derived the formula on the most general thermodynamical grounds [10]. But this is of minor importance for our arguments here.

In neither of the two papers in 1905 does Einstein refer to his predecessors and there is no bibliography. The old controversy concerning the possible influence of the famous Michelson-Morley experiment on ether drift on the genesis of Special Relativity has never been resolved satisfactorily (see, e.g. APS News, March 2004, pp 4-5 for a recent discussion of the subject). Einstein himself did not help the controversy to be resolved, adding from time to time new mysteries to the subject. It seems unlikely that the experimental result, even if it was well known at the time, could be crucial for postulating the central concept of the Special Relativity - the absolute speed of light. First of all, the result was neither the only one available and secondly, it was far from convincing considering the statistical nature of the method employed. Einstein resorted to an epistemologically decisive option. He turned to the most primitive experience (ontological view), but of a special kind - the gedanken experiment (epistemological view). It is ironical that he resorted to Newton's epistemology, to modify his basic notions of space and time.

Both results published in those papers have since been considered as Einstein's own contributions, stemming from his mind like Athens coming from Zeus' head. The reason for this was surely the fact that he offered a single underlying idea for both results; the concept of the extraordinary nature of light as a primitive construct. The lack of reference to other, previous or contemporary, authors might have been considered as risky, had not it concerned final results that were already known - Lorentz's transformations and $E = mc^2$.

## 6. Einstein and cosmogenesis

If both above mentioned papers dealt with subjects already in the air, the construction of the General Theory of Relativity has been considered as a great achievement of a single mind of genius. Though the motivation for the generalization of the physical situation from inertial to noninertial frames of reference looks straightforward, the task was too ambitious even for Einstein who lacked the necessary mathematical background for setting up the equation that was to replace Newton's dynamics. The story of devising the famous equation which connects three most fundamental physical quantities space, time and matter is well known.  Einstein acted as an inspired manager and creator until the equation appeared in its final form. (That Hilbert derived it about the same time as Einstein is of little importance for us here, though some authors refer to the equation as the Einstein-Hilbert expression.)  As an admirer of Mach's approach to mechanics, more precisely of his epistemology, Einstein was eager to incorporate Mach's idea that the local properties within a finite part of the Cosmos are determined by the overall influence from the rest of the Universe. In particular, Mach's principle, as Einstein termed it, that the inertia of a massive body depends on the mass distribution and the gravitational force of the Universe acting on the test object. It is this concept, that Einstein never incorporated fully into the theory, lead him in 1917 to apply the same theoretical construct to the Universe as whole. His model of a Cosmos without boundaries, a sort of closed infinity, was the first fully scientific, mathematically rigorously determined, Universe. With his model modern cosmology started its relentless advance.

What might be the feelings of this modern creator of the Cosmos when devising something that has always been the domain of divine? Interestingly, his model was a static one, a Cosmos without time and devoid of global evolution. Hence, Einstein could not be considered as a Demiurge, in Platonic terms, or the first Mover, as Aristotle termed it.  We shall come to this point later on.  Here we concentrate on the very notion of devising and describing the Universe and the possible psychological consequences for the human mind.

Einstein was the first to introduce mathematics into cosmology, but as for the physical aspects, Kant could be considered to be the first scientific cosmologist [11].  What were Kant's feelings while enquiring into the divine?  He was fully aware of his delicate position and protected himself from both possible forms of attack. In the actual dedication of his famous tract to the King, he apologized for his bold intrusion into the forbidden domain of the divine by expressing his awareness of his humble position. In the tract, he tried to protect himself from the inevitable assault by clerics by defending his concept of an infinite Universe through a reference to God's omnipotence. This tactic had some risk, as Galileo found when trying to interpret the Holy Scriptures to his advantage (or rather to Copernicus' advantage)[12]. By the time of Einstein however, European emancipation had moved from a theocracy to a secular society and contemporary cosmologists were not worried about a harsh clerical response and certainly not about an inquisition.  But a rational emancipation at the conscious level is only part of the story. The fear of the divine, deeply rooted in the human subconscious, acts as an archetypal barrier between the liberal mind and the traditional layers of dogma deposited through centuries, if not millennia. (The famous accident that Omar Khayam experienced after blasphemous shouting is a case in point).  It is this conflict between the rational and irrational that shaped Einstein's attitude towards the relation of faith to science, as we shall see in the following.

### 7. Einstein and microcosms

Although he did not produce either of the two formulations of Quantum Mechanics, in Einstein's contribution to the development of Wave Mechanics and subsequently to its interpretation, the epistemological background can not be overestimated. But as General Relativity ascribed to the previous theory of space and time the attribute 'Special Relativity', so with the advent of Quantum mechanics, both Heisenberg's Matrix Mechanics and Schr\"odinger's Wave Mechanics, the previous physics, relativistic and otherwise, became labelled classical theory. But despite his active involvement in the development of the new theory of matter, Einstein remained a classical physicist. The same thing happened to his generation as to the Pythagoreans who discovered the irrational number; its discovery destroyed the entire ideological base of their philosophy. The stochastic, intrinsic probabilistic nature of the new theory did not suit the classical mind, which experienced it as an epistemological failure to comprehend the complex nature of the microscopic world.  Interestingly the proverbial resistance of Einstein to the indeterministic interpretation of Quantum Physics came after his significant contribution in 1905 to the description of the epitome of stochastic behaviour - Brownian motion.  But here we are more interested in the psychological aspects of his assertion that Quantum Mechanics is an incomplete theoretical description of the microcosmical reality. We may speculate as to why he could not accept the probabilistic concept of the laws of nature, but here we just note that the motivation might stem either from epistemological or psychological sources (or possibly from both). From an epistemic viewpoint, the traditional wisdom was that probability comes in when the empirical evidence of the reality is deficient (ontological view).  Generally, the probabilistic approach is adopted when describing macroscopic effects whose sources are at the microscopic, inaccessible level.

The psychological resistance to the acceptance of uncontrollable events may be traced to a need to defend the power of the human mind to understand physical reality. Einstein was not the only one to express his skepticism concerning the completeness of the quantum mechanical description of reality (Schrödinger himself was one of the opponents to the Copenhagen interpretation of the wave function, though he subsequently complied with the general view), but his opinion on the matter carried particular weight, because of his reputation at the time. After the confirmation in 1919 of one of the principal effects predicted by General Relativity (the amount of light deflection in a gravitational field), the "suddenly famous Dr Einstein" (as a newspaper described him at the time) was considered the highest authority on the subject. The more so considering that he personally contributed to the formulation and rise of the new fundamental theory of microphysical reality. This was a manifestation of the attitude of a human mind with regard to its own abilities. But what about the divine? If the outcome of an experiment cannot be predicted, does it mean that it is inherently unpredictable to anybody, including God? And here we come to the crux of the matter, as is best illustrated the best by the famous Einstein-Bohr argument on the issue.

 To Einstein's assertion "God does not play dice", Bohr responded "Who are you to decide what God is supposed to do?" Both arguments expose nicely the dichotomy, which remained with Einstein in later life with regard to religion.

 Despite his humbly acknowledgement of his limitations, which we shall discuss later on, his human pride built upon his remarkable intellectual achievements could not be concealed from an attentive listener. His ambivalence toward the divine is even better expounded by his response to the (false) reports that ether was detected, by the famous phrase "Subtle is the Lord, but malicious he is not" [13]. It is difficult to escape the notion that Einstein, at least on this occasion, treated God as his partner, whose loyalty he found it necessary to defend. We shall return to this point later on, when discussing the parallel with Moses.

The issue of determinism versus indeterminism in the microscopic world revolves around the meaning of "determinable". It has at least three levels of meaning. Firstly the technical one connected to experimental feasibility. Secondly, the epistemic one, which is an intrinsic feature of a particular theoretical framework, Quantum Mechanics in this case. The most abstract meaning transcends the sphere of science and implies the absolute attribute of a physical reality, irrespective of human mental constructs, even of the most general theoretical type. It is here that the notion of divine power comes in. Einstein ultimately accepted that Quantum Mechanics could and should be considered complete, but that it did not mean that one might one day develop a more general theory that is deterministic. The question then arises as to whether God's interference with physical reality, whose creator he is supposed to be, is equivalent to our eventual possession of such an omnipotent theory. Or to put it another way, is it possible that a creator is not capable of controlling his creation? Or in view of Vico's argument mentioned above, that He does not comprehend his design? But is this a real issue at all? And here we come to the essence of religious versus rational thought.

Is there a genuine religious attitude bound to a rational mind, such as Einstein's? Einstein must have been be aware, at least subconsciously, that there is nothing and that there cannot be anything outside the human mind. One need not go back to Xenophanes and his famous dictum that it is not gods who created men, but the other way around. The issue that Einstein (and the rational mind in general) was facing is the same as the Eleatic philosophers put it - what are the human abilities concerning their own mental powers? [2] More precisely, can every gedanken problem, put forward by one human mind, be resolved by another human mind? Or more abstractly, in Gödel's sense, can we hope to conceive a reasonably general mental construct that is devoid of contradictions, paradoxes and conundrums? The issue is not one of confrontation between the human and the divine, but the completeness of human mentally constructed systems. It can also be considered according to Yung, as a tension between the archetypal and the rational [14]. The latter issue may be best epitomized by Yung's experience with Pauli's subconscious as revealed by his dreams.

## 8. Einstein and Judaism I.

Einstein's life and work was deeply connected to the historical development of Science, as conceived by the ancient Greeks and rediscovered by da Vinci, Galileo, Descartes, Newton and other European Hellenistic thinkers. On the other hand, he belonged to a small Jewish community, immersed in a large Christian European 'Sea'. In the above sections we have dealt with intrinsic features of the tension between the rational and the religious, as they emerged from these two principal pillars of European culture. We shall now consider a number of external factors, which determined Einstein's attitude towards religion.

Albert Einstein was born into a German Jewish family on March 14, 1879. His parents were not particularly religious and although they never rejected their Jewish faith they did not strictly follow the traditional rites and never attended religious services. However, when Einstein, at age six, entered a Catholic public primary school in Munich, they hired a tutor to teach him about Judaism in order to counteract his compulsory Catholic instruction. During that time he gained a deep religiosity and started to follow religious prescriptions in every detail. In his 1949 autobiography [15], Einstein states that his religious sentiment was originally initiated by traditional education. Nevertheless, the fact that he was even at such a young age strongly influenced by nature and music [16], obviously made him suitable material for the acceptance of religious ideas. To understand Einstein's religiosity one must bear in mind that this complex feeling emerged from an entangled mixture of nature, music and God [16]. Later on, close to his 13th birthday, he became completely irreligious and refused to go through with his *bar mitzvah,* but it seems that the feelings of reverence that he felt when in contact with nature were present all his life [17].

The origin of Einstein's conversion lies in the novel ideas that he acquired through reading scientific books. This led him to the conviction that the stories in the Bible could not be true and as he was an independent spirit he became suspicious of every kind of authority. Since there is but one step from denying authority to finding a replacement, we think that his attitude towards science and religion should be considered as starting from this point. It should be noted that Einstein did not attend religious services, nor did he pray at a place of worship of any kind. His civil marriage to Mileva Maric, who belonged to the Serbian Orthodox Church, also shows Einstein's indifference towards religion affiliations. On the other hand, despite his refutation of Orthodox Judaism, he saw himself as a Jew. In his interview with Peter Bucky we find following statement through which he tried to clarify his position [18]: "Actually it is a very difficult thing to even define a Jew. The closest that I can come to describing it is to ask you to visualize a snail. A snail that you see near the ocean consists of a body that is snuggled inside a house which is always carried around. But let us picture what would happen if we lifted the shell off the snail. Would we not still describe the unprotected body as a snail? In just the same way, a Jew who sheds his faith along the way, or who even picks up a different one, is still a Jew."

## 9. Einstein and Judaism II.

> *Had the good Lord consulted me while creating the World,*
> *I could have given him some good advice.*

> Alphonso X

As his fame grew, the number of Einstein's texts concerning science and religion gradually increased. In his writings and interviews

Einstein's statements are sometimes ambiguous, even contradictory, but it is easy to recognize some key facts in his opinions.  Einstein's starting point was refutation of the traditional concept of a personal God, a God who rewards and punishes the object of His creation: "I cannot then believe in this concept of an anthropomorphic God who has the powers of interfering with these natural laws.... If there is any such concept as a God, it is a subtle spirit, not an image of a man that so many have fixed in their minds" [18]. In his reply to one of the letters sent to him in Princeton he was even more explicit: "I do not believe in a personal God and I have never denied this but have expressed it clearly" [17]. On the other hand, after refusing to implement purpose, goal or anthropomorphic principle into Nature, Einstein introduced the notion of "cosmic religious feeling" through which he tried to summarize his beliefs [19].  Basically, cosmic religious feeling concerns his belief in the rational structure of the world. By entering into the field of science, we are trying to grasp the "grandeur of reason incarnate in the existence which, in its profound depths, is inaccessible to man" [20]. This leads to a mysterious experience, which arises from an awareness of the insufficiency of the human mind to fully understand the harmony of the Universe and is the core of Einstein's religious feeling [21]. Although, throughout these debates, Einstein tried to keep an autonomous position ("I'm not an atheist, and I don't think I can call myself a pantheist" [16]), his religious views can be considered pantheistic. Some remarks about Spinoza; "I believe in Spinoza's God who reveals Himself in the orderly harmony of what exists, not in a God who concerns himself with the fates and actions of human beings"[22], or on Buddhism; "The religion of the future will be cosmic religion, the religion which is based on experience and which rejects dogmatism. If there's any religion that could cope with scientific needs it will be Buddhism...." [17] can easily support the former conclusion.

How can one elucidate the underlying rationality of the Universe? Einstein directs us to mathematics: "Our experience hitherto justifies us in believing that nature is the realization of the simplest conceivable mathematical ideas. I am convinced that by means of purely mathematical constructions we can discover the concepts and the laws connecting them with each other which furnish the key to the understanding of natural phenomena.  Experience may suggest the appropriate mathematical concepts, but they most certainly cannot be deduced from it. Experience remains, of course, the sole criterion of the physical utility of a mathematical construction. But the creative principle resides in mathematics. In a certain sense, therefore, I hold it true that pure thought can grasp reality, as the ancients dreamed" [23]." The origin of these ideas can be traced to Kant's "Critique of pure reason". Einstein was a serious reader of Kant's philosophy.  Besides scientific books, he had read the "Critique of pure reason" at an early age, just before he refused to take the the the bar mitzvah [16]. Kant attempts to explain in his book how mathematics is possible in the first place [24]. We can treat mathematical knowledge in two ways: as empirical in its essence, which is essentially Hume's viewpoint, or as an outcome of pure reason which is Descartes' viewpoint [24]. In the first approach the *a priori* truthfulness of mathematics is just an image through which nature rescues us from the lack of pure reason and if we accept Descartes' view, we must explain why this invention of our spirit is so successful in practice? To answer this question one might obviously use an ontological argument [24]. For that reason Kant turns to the notion of subject, which becomes a crucial point of his philosophy. He started with the well-known fact that for each subject the objects of the outside world are actually mental representations or phenomena. This does not mean that he took the solipsistic position and denied the existence of objects outside our senses (*Ding an sich*). Kant just tried to make the difference between the thing as it is and the thing as we know it, a phenomenon. On the other hand, this introduces the problem of establishing something common to the mental representations of all subjects, something that can be called knowledge. Kant attempted to solve this problem (i.e. gaining knowledge and/or the existence of mathematics) by introducing the notion of an *a priori* intuition of time and space. However, his solution first induced reactions of romantic idealism and later on of the other schools of philosophy. Since this subject is still a matter of dispute, we will leave it aside and concentrate on Einstein's approach.

Obviously, Einstein assumed that mathematics can offer us knowledge about the Universe or, following the above discussions, a perspective, that is independent from our mental representations. His view is Platonic. It can be understood as a combination of the Platonic school of mathematics, which claims that mathematical objects are not derived but possess an autonomous existence, and the opinion that they (i.e. these mathematical objects) can be directly realized in nature. The discovery of the hidden rational nature of reality should be the principal goal of humankind, as he pointed out at the end of his article from the Symposium on Science, Philosophy and Religion in New York 1941. [20]: "The further spiritual evolution of mankind advances, the more certain it seems to me that the path of genuine religiosity does not lie through fear of life or death and blind faith, but through striving after rational knowledge. In this sense I believe that the priest must become a teacher if he wishes to do justice to his lofty educational mission." Furthermore, Einstein considered the people who acted according to these principles as the "priests" of his religion. "The religious geniuses of all ages have been distinguished by this kind of religious feeling, which knows no dogma and no God conceived in man's image; so that there can be no church whose central teachings are based on it. Hence it is precisely among the heretics of every age that we find men who were filled with this highest kind of religious feeling and were in many cases regarded by their contemporaries as atheists, sometimes also as saints. Looked at in this light, men like Democritus, Francis of Assisi, and Spinoza are closely akin to one another [19]." For this reason, it is not unusual that he himself took on the task of continuing the endeavours of these great people from the past. As a voice of novel spiritualization, he started to play the role of Moses, a prophet of the new faith manifested through cosmic religious feeling. As was mentioned earlier, a closer look at the history of religion and philosophy reveals that ideas about exceptional religious personalities are almost permanently present in the western or Judeo-Christian civilization. Besides an obvious influence of the Torah, with prophet figures so immanent to Judaism, we can also find spiritual heresies present in Christianity since the time of Joachim of Fiore (ca.1132-1202). In these heresies (which sometimes include the work of philosophers, Hegel for example) the teachings about the Holy Spirit are emphasized to such an extent that even incarnation, the personification of God, becomes a continuous and at all times a present and repeatable event [25]. Bearing this in mind, it can be argued that Einstein's position as a prophet is not completely unjustified. If we adopt the above definition of knowledge as something common to our individual, phenomenological experiences, then Einstein has indeed created our world. His General Relativity theory gives us knowledge about the Universe, a picture of the world that exists independently of our senses, that is in fact, the maximum that we can grasp with our feeble minds. Therefore, we might still consider Einstein as a Demiurge, a God creator. The question remains, of course, whether he was aware of this position and whether he played on this, consciously or unconsciously. He wrote [26] "When I am judging a theory I ask myself whether, if I were God, I would have arranged the world in such a way." This was not merely a repetition of the famous remark by Alphonso the Wise, since the Castilian king was a mere organizer of a compilation of astronomical tables, while Einstein was devising theoretical models, which could reflect the physical reality itself.

The reverence which his eminent colleagues felt with regard to Einstein's scientific achievements surely added additional weight to his feelings of self-respect. Here it is what Paterniti wrote in his book [27].

'Another contemporary of Einstein, Erwin Schrödinger, claimed that Einstein's theory of relativity quite simply meant "the dethronement of time as a rigid tyrant", opening up the possibility that there might be an alternative master plan. "And this thought", he wrote, "is a religious thought, nay I should call it the religious thought." With relativity, Einstein, the original cosmic slacker, was himself touching the mind of a new god, trying to wriggle through some wrinkle in time. "It is quite possible that we can do greater things than Jesus," he said'.

The dethronement of time, with the latter being the most fundamental and elusive entity within the physical world, meant at the same time "overruling" the most reverent Hellenic god, Chronos. When Kurt Gödel finds in 1949 that Einstein's General Relativity allows for the so-called time-like curves, the "rigid tyrant" was not only overthrown, but killed altogether. (It turns out that Alice's cry in Wonderland "He is killing time!" was a prophetic warning to unrestrained scientific speculations.)

The last sentence in the above quotation was an obvious allusion to Jesus' "tunneling" through the "spatio-temporal barrier" between the Crucifixion and Resurrection, and the so-called "wormhole" in the four-dimensional space-time manifold (the ideological background of the modern time machines). The mild irony, so characteristically present whenever Einstein referred, albeit indirectly, to religion reminds us of his ambiguous attitude to the subject. We shall return to this sentence later on, when discussing his relation with Christianity.

Apart from these analogies, we can not overlook Einstein's manners and behaviour in his old age. His unorthodox clothing, avoidance of sockets, using rope instead of a belt, his general appearance resembling that of Chaplin, all this points towards the lifestyle of a hermit. True, this could be equally interpreted as a disregard of petit bourgeois conformism, the latter being so far from his non-conformism in every respect. But one may equally assign it to his prophetic self-image, more precisely, to a subconscious awareness of being a law-giver. The latter is particularly indicated by Einstein's prominent hair, which inevitably reminds one of Samson and other biblical symbols of might (which, in its turn, was borrowed from the paradigm of the lion's mane). Moreover, if we recall the death of Moses in the way Bible presents it: "So Moses the servant of the Lord died there in the land of Moab, according to the word of the Lord, and he buried him in the valley in the land of Moab opposite Beth-pe'or; but no man knows the place of his burial to this day" (Deuteronomy 34:5-6), a strange parallel to the destiny of Einstein's whereabouts arises. After his death Einstein was cremated according to his wishes and his ashes were scattered at a secret spot on the Delaware River (probably into water) [27]. It seems that he was ready to play the role of his great ancestor to the very end. (The analogy between the cases of Mileva Maric and Agar (Hagar) comes to mind, too, but we shall not dwell on it here).

Another, equally valid, interpretation would be to conceive the act as a religious ritual of unifying with the Nature-God. The choice of water is indicative as a generic dissolver in many religions. It is a primeval element as well, as with Thales. It is also the substance that washes our human sins, albeit in a symbolic manner, and is present in numerous religious movements, such as the Essenes [1], Hinduism and Christianity for example. The choice of river waters is not insignificant, too, for the water flow epitomizes the everlasting change in Heraclites' sense, and reminds us of the transient nature of our life. That all these motives appear disjunctive if not contradictory when taken altogether, should not bewilder us, since the very concept (and phenomenon) of death is counter-intuitive in itself.

It is also interesting to comment on Einstein's attitude towards political power. His refusal to take the post of the President of Israel may be interpreted in many ways, but we shall consider here only one aspect which one might term a prophetic one. It is a well known Old Testament tradition that prophets considered themselves as messengers of God and never engaged in fighting for power, in particular for the position of a sovereign. This was for good reasons. Firstly, they did not have to share political responsibilities if things turned out badly, while retaining their right to criticize the government (a position which modern heads of churches hold until now). Secondly, they are protected in their activities by the Supreme Being, and thus keep their august position relative to earthly power [28]. Thirdly, they retain their independence from the mob, for ruling implies a mutual dependence between the dominating and dominated. Last, but not least, these wise men knew that it is the balance between spiritual and political power that keeps a society stable and functioning. Einstein used to distance himself from his environment, both family and social, (even scientific), since his fame acquired global dimensions. Engaging in any sort of official public activity would break his "splendid isolation" and would surely spoil his self-image of somebody who is "above everything".

    We now come to the last point of our discourse.

## 10. Einstein and Christianity

So far we have been dealing with Einstein's link to the Old Testament tradition. We shall finish our analysis of his religious attitude by considering briefly his relations with the faith that was prevailing in his immediate environment, Christianity. The latter has been involved in Einstein's development regarding religion in many ways, albeit implicitly. It helped a young schoolboy, while attending parallel religious lessons in Mosaic and Christian dogma, to realize the naivety of religion. At a more mundane level, early Christian tradition in its ascetic aspects surely did not fail to influence Einstein's manners, both concerning his way of life and his attitude towards the fame that he was the object of in the second half of his life. These manners oscillated between subtle arrogance and humble modesty, just as Christ's attitude towards the environment used to jump from servitude to warnings of the wrath of God. One is tempted to doubt that his modesty was sincere, but there is no reason to disbelieve Einstein's honesty. After all, was it not the humiliation syndrome that secured both the Jewish survival and the victorious Christian march through numerous persecutions?

One might object that this interpretation is redundant in view of what has been said about Einstein's prophecies, but we know that the Christian way of gaining dominance was also achieved according to the Old Testament religious strategy.

Concerning the actual personality of Jesus of Nazareth, Einstein shared the common attitude of his "apatrid compatriots". As he confessed himself, he did not doubt Jesus' existence as a real man whose life was so vividly described in Gospels. As for Jesus' miracles, divinity and other religious and mythical attributes, he did not have to be particularly explicit, in view of his general attitude concerning the concept of a personal god. He occasionally made allusions to the matter, as in the aforementioned quotation on Jesus' achievements according to New Testament.

As somebody who was not committed to any particular religious faith, being aware of his Jewish heritage and at the same time imbedded into quasi-Judaic Christian 'Sea', Einstein was occasionally very insensitive in dealing with intimately delicate and socially risky situations. In his letters to his future wife, Else, he used to refer to his current wife, Mileva, by marking her simply with a cross and omitting her name. Whether he referred to Mileva's Serbian-Orthodox origin, or whether it was a mere allusion to her as a "burden" in his life (Albert was about to divorce Mileva), or both, is a matter of choice (or taste).

Talking about his first wife Mileva Maric, it is interesting to quote a passage from [9], related to Albert's and Mileva's stay at her father's cottage at Kach in Vojvodina, then in the Austro-Hungarian empire. (It was just after submitting his "Special Relativity" paper to Annalen der Physik, and Mileva told proudly her parents that the paper will make her husband famous).

"[He] liked the most riding a donkey. He noticed soon that wherever he rode a flock of sheep used to follow him, even when they were rather far away. He asked Rada if it was he who trained the sheep to follow the donkey, but the answer was negative. Albert was quite curious about the behaviour of the sheep and started analyzing the influence of the mutual distance and velocity between the donkey and sheep on the "attraction" of the latter. Workers on the estate found Albert's "investigations" quite odd and used to refer to him as 'that crazy Maric's son-in-law'".

One could not help thinking, reading this passage, that if somebody wanted to arrange an allegory on the famous New Testament episode of Jesus entering Jerusalem (or even on the entire New Testament mythology), he could not have done it more picturesquely.

On a more ideological level, one is tempted to see in Einstein's insistence on an impersonal God with his secrets imbedded in Nature to be revealed by human mind, a kind of gnosis. Though the latter was in all probability developed somewhat before Christian faith appeared on the religious stage, it has always been linked with Christianity, as one of its specific heresies [14]. But surely Einstein was not a mystic and in his case one might rather think in terms of Spinozian pantheism with an overlay of scientific curiosity. On the other hand Einstein never considered scientific inquiries as a natural human impulse, devoid of any external or internal motivation. In this respect he lacked the Hellenistic (hedonistic) experience of intellectual activity as an autonomous human need. At least he did not show it in his personal communications with the environment.

## 11. Epilogue

Einstein was not a religious thinker and it would not do justice to him to judge his opinion and attitude toward religion in terms of self-consistency, or even intellectual evolution. But one can certainly discern in his addressing the issue an endeavour to fuse two principal sectors of human life, religious (irrational) and scientific (rational) ones. Though he never said it explicitly, his motivation was to formulate a unique point where these two aspects merge again, having been separated in archaic times as the Biblical narrative on the Original sin informs us, albeit in allegorical terms. Just as he was striving during the second part of his life to formulate what one would term today as 'The Theory of Everything', by trying to fuse his theory of gravity with Maxwell's electrodynamics.

Einstein was not an explicit atheist, unlike Marx and Freud for example, but neither was he an orthodox believer. He failed to unify gravity and electromagnetism, since in all probability they are incompatible in the sense that gravity is not a force at all (as Einstein believed himself). Likewise, to know and to believe will most probably remain forever separate areas of human intellectual activity. Even as powerful a mind as that of Einstein could not overcome this dichotomy.

.

**Acknowledgements**. We are gratefull to Dr Gillian Peach for her critical reading of the manuscript and numereous useful sugestions. We acknowledge with pleasure useful conversations with Dr Zoran Stokic. We also wish to thank Milica Manojlovic for her assistance in the literature search. This work has been partially supported by the Ministry of Science and Environment Protection of Republic of Serbia.

*Vladimir Djoković*
*Institute for Nuclear Sciences "Vinca",*
*P.O. Box 522*
*11001 Belgrade,*
*Serbia*
*e-mail: djokovic@vin.bg.ac.yu*

*Petar Grujić*
*Institute of Physics*
*P.O. Box 57*
*11080 Belgrade*
*Serbia*
*e-mail: grujic@phy.bg.ac.yu*